\documentclass[12pt,journal,compsoc]{IEEEtran}
\ifCLASSOPTIONcompsoc
\else
\fi
\usepackage[cmex10]{amsmath}
\usepackage{epsfig,amsmath,amssymb,graphics,graphicx,color,calc,cite,psfrag,cases}

\def\0{{\tt 0}}
\def\1{{\tt 1}}
\def\identity{ {1 \mskip -5mu {\rm I}}  }

\let\id=\identity

\begin{document}
\newtheorem{theorem}{Theorem}
\newtheorem{Lemma}{Lemma}
\newtheorem{Proposition}{Proposition}
\newtheorem{Corollary}{Corollary}
\newtheorem{remark}{Remark}
\newtheorem{claim}{Claim}

\title{Group Testing with Random Pools: \\ optimal two-stage algorithms }

\author{Marc M\'ezard, Cristina Toninelli
\IEEEcompsocitemizethanks{\IEEEcompsocthanksitem
M.M\'ezard is with CNRS, LPTMS,
Batiment 100, Universit\'e de Paris Sud
91405 Orsay - France (e-mail: mezard@lptms.u-psud.fr);
\IEEEcompsocthanksitem
C.Toninelli is with  CNRS, LPMA, Boite courrier 188, 4 Pl. Jussieu,
Univ.Paris VI-VII, France (e-mail: ctoninel@ccr.jussieu.fr)}
\thanks{This work has been supported in part by the EC grant ``Evergrow'', IP 1935 of FET-IST.}}

\markboth{}
%Journal of \LaTeX\ Class Files,~Vol.~6, No.~1, January~2007}
{M\'ezard and Toninelli: Group Testing with Random Pools}
\IEEEcompsoctitleabstractindextext{%
\begin{abstract}
%\boldmath
  We study Probabilistic Group Testing of a set of $N$ items each of which is
  defective with probability $p$. We focus on the double limit of small
  defect probability, $p\ll 1$, and large number of variables, $N\gg 1$ ,
  taking either $p\to 0$ after $N\to\infty$ or $p=1/N^{\beta}$ with
  $\beta\in(0,1/2)$. In both settings the optimal number of tests which are
  required to identify with certainty the defectives via a two-stage procedure,
  $\overline T(N,p)$, is known to scale as $Np|\log p|$. Here we determine the
  sharp asymptotic value of $\overline T(N,p)/(Np|\log p|)$ and construct a
  class of two-stage algorithms over which this optimal value is attained.
  This is done by choosing a proper bipartite 
  regular graph (of tests
  and variable nodes) for the first stage of the detection. Furthermore we 
  prove that
  this optimal value is also attained on average over a random bipartite graph
  where all variables have the same degree, while the tests have
  Poisson-distributed degrees. Finally, we improve the existing
  upper and lower bound for the optimal number of tests in the case
  $p=1/N^{\beta}$ with $\beta\in[1/2,1)$.
\end{abstract}
\begin{IEEEkeywords}
Group testing, reconstruction algorithms
\end{IEEEkeywords}}

\maketitle

% is not selected <OR> if conference mode is selected - because compsoc
% conference papers position the abstract like regular (non-compsoc)
\IEEEdisplaynotcompsoctitleabstractindextext

% \fi
\IEEEpeerreviewmaketitle

%\setlength{\textwidth}{170truemm}
%\begin{document}
%\title{Group testing with Random Pools: Phase Transitions and Optimal Strategy}
%\author{MM+CT}
%\date{\today}
%\maketitle
\section{Introduction}

The aim of {\sl Group Testing} is to detect an unknown subset of {\sl
  defective} (also referred to as positive or active) {\sl items} out of a set
of objects by means of queries (the tests) in the most efficient way. In other
words we are given a set of objects, ${\cal{O}}$, which contains an unknown
subset of defectives, ${\cal{D}}$, and the task is to identify ${\cal{D}}$ by
means of the fewest possible number of tests. Tests are
queries of the form ``Does the {\sl pool} ${\cal{Q}}$ (where ${\cal{Q}}$ is a
subset of ${\cal{O}}$) contain at least one positive item?''.
This problem was originally introduced in relation with efficient mass blood
testing \cite{Dorfman}. Afterwards, it has been also applied in a variety of
situations in molecular biology: blood screening for HIV tests \cite{Zenios},
screening of clone libraries \cite{clone1,clone2}, sequencing by
hybridization \cite{sequencing1,sequencing2}, \dots. Furthermore it has proved
relevant for fields other than biology including quality control in product
testing \cite{control}, searching files in storage systems \cite{searching},
data compression \cite{compression} and more recently in the context of data
gathering in sensor networks \cite{sensor}. We refer to
\cite{book,review1} for reviews on the different applications of GT.
Here we will deal with  the very much studied {\sl gold-standard} case, namely
the idealized situation in which tests are perfect: there can be neither false positives nor
false negatives in the test answers. It is important to keep in mind for future work that,
however, in many biological applications one should include the possibility of
errors in the test answers.

Before presenting our results we recall some standard
classifications of GT problems.  First of all
a GT problem can be either {\sl Combinatorial} or {\sl Probabilistic}.
Combinatorial GT refers to the
situation in which $\cal D$ can be any member of a predetermined class
of sets in $\cal O$. The task is here to find the algorithm which
requires the minimal number of tests to determine $\cal D$ in the
worst case.  In {\sl probabilistic GT} we
are given a configuration space $S$ and a probability distribution
$\mu$ on $S$ and the set of objects $\cal{O}$ (and therefore the
corresponding $\cal D$) is chosen in $S$ according to $\mu$.
In this case the
task is to optimize the expected (with respect to $\mu$)
number of tests required to determine
$\cal D$.
Furthermore in both combinatorial and probabilistic GT
there is an additional classification which concerns the
number of stages, i.e. parallel queries, in the detection procedure.  For
one-stage (or {\sl fully non-adaptive}) algorithms 
 all tests are specified in advance: the choice of
the pools $\{{\cal{Q}}\}$ does not depend on the outcome of the
tests (and therefore does not depend on $\cal O$). For several
 biological applications a non-adaptive procedure 
would in principle be the best one. Indeed the test procedure 
can be destructive for the objects and repeated tests on the same
sample require more sophisticated techniques.  However
the number of tests required by
fully non-adaptive algorithms can be 
much larger than for adaptive ones. The
best compromise for most screening procedures \cite{Knill}
is therefore to consider {\sl two-stage} algorithms 
with a first stage containing a set of
predetermined pools (tested in parallel) and a second stage whose
pools are chosen depending on the outcomes of the first
stage  (and therefore on the choice
 of $\cal{O}$). For Probabilistic GT
the only possibility to detect
all defectives with such a procedure is to choose
a {\sl trivial two-stage algorithms} \cite{Knill} which
individually tests on the second stage all the variables 
which are left undetermined by the first stage. Here we will consider
Probabilistic Group Testing when $\mu$ is
Bernoulli product measure and we will analyze
the performance of two stage algorithms
as a function of the overall number of objects, $N$,
and of the probability that a chosen object is defective, $p$. 
 In particular we will analyze the relevant limit of small $p$ and large $N$, which has already been investigated in (\cite{Berger,Vaccaro,Berger2,Knill}). 
A detailed account of our new contributions follows.

\section{Notation and results}
\label{results}

We consider {\sl Probabilistic Group Testing} in the {\sl Bernoulli
  p-scheme}: the configuration space is $S=\{0,1\}^N$, namely the set of all
vectors $X=(x_1,\dots,x_N)$ with $x_i=\{0,1\}$, and the probability measure is
Bernoulli product measure $\mu_p$ with marginal $\mu_p(x_i=1)=p$, namely
$\mu_p(X)=\prod_{i=1}^Np^{x_i}(1-p)^{1-x_i}$.
For a given choice of $X$ we say that {\sl variable $i$ is (is not) defective or positive}
if $x_i=1$ ($x_i=0$).

A test of the type ``Does pool ${\cal{Q}}$ contain at least a defective?''
corresponds here to asking whether the value of the random variable
constructed as an OR function among the variables of the pool equals one or
zero. More precisely we will call ``{\sl pool} $a$'' an $N$ component binary
vector $P_a=(c_{1a},c_{2a},\dots,c_{Na})$ with $c_{i,a}\in\{0,1\}$ and we will
say that {\sl variable i belongs (does not belong) to pool} $a$ if $c_{i,a}=1$
($c_{i,a}=0$). With this notation we will call ``{\sl test} $a$'' the random
variable $T_a\in\{0,1\}$ with $T_a=0$ if $c_{i,a}x_i=0$ for all
$i=(1,\dots,N)$, $T_a=1$ otherwise. In other words $T_a$ is the OR function
among the variables that belong to pool $a$.

For a given choice of the variables, $X$, and a set of $M$ pools, $\{P_a\}$, $a=1,\dots M$,
we say that: \\
(a) variable  $i$ is a {\sl sure zero} if
there exists at least one $a\in(1,\dots, M)$ such that: $i$ belongs to pool $a$ and $T_a=0$\
(b) variable $i$ is a {\sl sure one} if
there exists at least one $a\in(1,\dots, M)$ such that:
$i$ belongs to pool $a$, $T_a=1$ and all the other variables $j$, $j\neq i$,
which belong to pool $a$ are  sure zeros.\\ 
It is obvious that if $i$ is a
sure zero then $x_i=0$ and if $i$ is a sure one then $x_i=1$. Note that however
the converse is not true: there can be a zero (one) variable that is not a sure zero (sure one, respectively). Indeed, any given choice of $X$, $M$ and 
$\{{P}_a\}$ $a=(1,\dots,M)$, identifies the following subsets of 
${\cal V}:=\{1,\dots, N\}$.\\
(i) The zeros and the ones
$${\cal Z}:=(i: i\in {\cal V};~ x_i=0)$$
$${\cal D}:=(i: i\in {\cal V};~ x_i=1)={\cal V}\setminus \cal Z;$$
(i) The sure zeros and the sure ones
$${\cal Z}\supset{\cal {S}}_0:=(i: i\in {\cal V};~ \prod_{a=1}^M {\left(T_a\right)}^{c_{i,a}}=0)$$ 
%DC
%ancien: $${\cal D}\supset{\cal {S}}_1:=
%(i:i\in {\cal V};~ \sum_{a=1}^M c_{i,a}x_i\prod_{j\neq i}\identity_{{\cal S}_0}(j)>0)$$
$${\cal D}\supset{\cal {S}}_1:=
(i:i\in {\cal V};~ \sum_{a=1}^M c_{i,a}x_i\prod_{j\neq i}\left(\identity_{{\cal S}_0}(j)\right)^{c_{j,a}}>0)$$
(here and throughout this paper we define $0^0=1$ and
$\identity_A$ stands for the characteristic function of set $A$);.\\
(iii) The undetermined zeros and the undetermined ones
$${\cal U}_0:={\cal Z}\setminus {\cal {S}}_0, ~~~ {\cal U}_1:={\cal D}\setminus {\cal {S}}_1.$$
The two-stage algorithms that we consider are composed by a first stage of
parallel tests and a second stage of individual tests over the variables whose
value has been left undetermined by the first stage. Therefore the choice of
the algorithm is completely defined by fixing the number of tests in the first
stage, $M\in\mathbb N$, and by choosing the pools $\{P_a\}$, $a=1,\dots,M$. The
latter corresponds to fixing an $N\times M$ matrix $C_{N,M}$ with binary
entries $c_{i,a}\in\{0,1\}$ which give the i-th component of vector $P_a$.
This will be called the {\sl connectivity matrix}. In other words, the choice
of the algorithm corresponds to fixing a couple $(M,C_{N,M})$, namely choosing
a {\sl bipartite graph} 
$G={\mathcal G}(C_{N,M})$ with $N$ {\sl variable nodes} and $M$ {\sl test
  nodes}. The number of tests required to identify the defectives (i.e. to
decode the value of $X$), $T(X,M,C_{N,M})$, is therefore given by the number
of tests in the first stage plus the number of variables which are left
undetermined by them, namely
\begin{equation}
T(X,M,C_{NM}):=M+|{\cal U}_0|+|{\cal U}_1|.
\end{equation}
Note  that ${\cal U}_0$ and ${\cal U}_1$ depend
in general on $X,M$ and $C_{N,M}$. We will denote by
$T_{M,C_{N,M},p}$
the mean of $T(X,M,C_{N,M})$ over the Bernoulli distribution $\mu_p$  for $X$, namely

\begin{equation}
\label{meantest}
T_{M,C_{N,M},p}:=M+\sum_{X\in S} \mu_p(X) \left( |{\cal U}_0|+|{\cal U}_1|\right).
\end{equation}

In this probabilistic setting the first important issue
is to determine the optimal value
$\overline{T}(N,p)$ of $T_{M,C_{N,M},p}$ over all two-stage algorithms, i.e.
over all choices of $M$ and $C_{N,M}$
\begin{equation}
\label{optimalmeantest}
\overline{T}(N,p):=\min_{M,C_{N,M}}T_{M,C_{N,M},p}
\end{equation}
%M I tend to find it superfluous, mayb we could just say
%CR I agree, just changed $\min$ into optimal value in your sentence
where minimization is restricted to $M=(1,\dots,N)$ (it is obvious that
the 
%$\min$ 
optimal value can never be attained at $M\ge N+1$). 

Here we will study this problem in the relevant limit of small defective
probability, $p\ll 1$, which has already been investigated in
\cite{Berger,Vaccaro,Berger2,Knill}. 
We will denote by $\lim_{N \to \infty \vert \beta}$ the limit where
$N$ goes to $\infty$, $p$ goes to zero, with $p=N^{-\beta}$ and $\beta>0$, i.e.
\begin{equation}
\lim_{N \to \infty \vert \beta} f(N,p):=\lim_{N\to\infty}f(N,N^{-\beta}).
\end{equation}
We will also study the limit $\lim_{p\to 0} \lim_{N\to\infty}$ and, 
in order to
lighten the presentation of our results,
we will refer to this case as the $\beta=0$
case:

\begin{equation}
 \lim_{N \to \infty \vert 0}f(N,p) :=  \lim_{p\to 0} \lim_{N\to\infty}f(N,p).
\end{equation}

Our main contributions are the following
results for the asymptotics of $\overline{T}(N,p)$, which will be proved
in Section \ref{regreg} and \ref{upper}, respectively.\\

\begin{theorem}
\label{teo:betas}
When $\beta\in [0, 1/2)$, 
\begin{equation}
\label{asymptoticbsmall}
\lim_{N\to\infty\vert \beta}\frac{\overline{T}(N,p)}{N p |\log p|}=\frac{1}{{(\log 2)}^2}. 
\end{equation}
\end{theorem}

\vspace{0.2 cm}

\begin{theorem}
\label{teo:betal}
When $\beta\geq 1/2$, 
\begin{equation}
\label{asymptoticblarge}
 \frac{1}{{(\log 2)}^2}\leq \lim_{N\to\infty\vert \beta}\frac{\overline{T}(N,p)}{N p |\log p| }\leq  e.
\end{equation}
\end{theorem}

\vspace{0.2 cm}

To our knowledge the best previously known bound  for  
$0<\beta<1$ were 
\begin{equation}
\label{bergerbound}
\frac{1}{\log 2} \leq \lim_{N\to\infty\vert \beta}
\frac{\overline{T}(N,p)} {N p |\log p|}\leq \frac{4}{\beta}
\end{equation}
which have been obtained in \cite{Berger}:
the lower bound via the information theoretic bound and the upper bound by the
explicit construction of a decoding algorithm based on a random choice of the
pools.

Our results determine the sharp asymptotics of $\overline T(N,p)/(Np|\log p|)$
for the cases $p={N^{-\beta}}$ with $\beta\in (0,1/2)$ 
and for the cases $p\to 0$ after $N \to \infty$. Furthermore, they sharpen the previously existing bounds for $p={N^{-\beta}}$ with $1/2\leq\beta<1$. \\ 

A second relevant issue is the explicit construction of
an {\sl asymptotically optimal algorithm},
namely the identification
of a family of
couples  $(M,C_{N,M})$ such that 
\begin{equation}
\label{optimal}
\lim_{N\to\infty\vert \beta}\frac{T_{M,C_{N,M},p}}{Np|\log p|}=
\lim_{N\to\infty\vert \beta}\frac{\overline{T}(N,p)}{Np|\log p|}.
\end{equation}
Here, for each $\beta$ with $0\leq\beta<1/2$, we identify  
a ($\beta$-dependent) family of couples $(M,C_{N,M})$ 
which satisfies (\ref{optimal}).
Let
\begin{equation}
\label{l}
\overline L=:[|\log p|/\log 2]
\end{equation}
\begin{equation}
\label{m}
\overline M=:[Np|\log p|/(\log 2)^2]
\end{equation}
 where $[x]$ stands for the integer part of $x$. 
We construct a pooling design based on a 
``{\it regular-regular }'' bipartite graph
with $N$ variable nodes, $\overline M$ test nodes,
$\overline L$ tests per variable and $ \overline K=N \overline L/ \overline M$ variables per tests, and with girth (i.e. length of the shortest graph cycle) at least $6$.
 This means that the corresponding
 connectivity matrix satisfies the following 
 constraints

\begin{IEEEeqnarray}{c}
\label{condition1}
\forall a\in(1,\dots,M): \ \sum_{j=1}^N c_{j,a}= \overline K \\
\label{condition2}
\forall i\in (1,\dots, N):\ \sum_{b=1}^{ M} c_{i,b}\!=\!\overline L~\\
\sum_{1\le j<l\le N}\sum_{1\le d<b\le M}c_{j,b}c_{j,d}c_{l,b}c_{l,d}=0
\label{conditionmatrix}
\end{IEEEeqnarray}
In Section \ref{regreg} we will prove that
any such connectivity matrix is
asymptotically optimal, namely

\vspace{0.2 cm}

\begin{theorem}
\label{teo:regbetas}
Let $C_{N,\overline M}^{\overline L}$ be such that conditions
(\ref{condition1}), 
%DR added
(\ref{condition2}) and (\ref{conditionmatrix}) are satisfied.
If $0\le \beta<1/2$, then:
\begin{equation}
\lim_{N\to\infty\vert \beta}
\frac{T_{\overline {M},C_{N,\overline {M}}^{\overline L},p}}{N p |\log p| }= \frac{1}{{(\log 2)}^2}.
\end{equation}
\end{theorem}

\vspace{0.2 cm}

Notice that 
the family of graphs satisfying the requested properties is non-empty under
the conditions for $\beta$ stated in the theorem,
thanks to a constructive procedure found in \cite{LuMoura}, as we shall
discuss in the proof.

\vspace{0.1 cm}

Furthermore we have proved that the optimal value is also attained
asymptotically by some random pool designs whose construction is much simpler
than the one 
of  \cite{LuMoura}.
Let $P^{{\cal{R}}-{\cal{P}}}_{N,M,L}$ denote the distribution
 of bipartite ``{\it regular-Poisson}'' graphs 
with
$N$ variable nodes, $M$ test nodes, and 
 a fixed number  of tests,  $L$, randomly connected to each variable node. Explicitly: 
 \begin{equation}
\label{regpoi}
 P^{{\cal{R}}-{\cal{P}}}_{N,M,L}(C_{N,M}):=\prod_{i=1}^N P(c_{i,1},\dots,c_{i,M})
\end{equation}
with 
 \begin{numcases}{P(c_{i,1},\dots,c_{i,M}) :=}
 \nonumber
 \frac{1}{\binom{M}{L}} & {\mbox{if}} $\displaystyle{\sum_{a=1}^{M} c_{i,a}=L}$\\
 0 & {\mbox{otherwise.}}
     \end{numcases}
 Note that, when one takes the large $N$ limit
with  $L\ll N$ and $L\ll M$,  the degrees of the tests become iid random variables 
with a Poisson distribution of mean 
 $K=NL/M$. If we make the choice $L=\overline L$ and $M=\overline M$ as in (\ref{l}) and (\ref{m}) the following result, whose proof is provided in section \ref{upper}, holds

\vspace{0.2 cm}

 \begin{theorem}
 \label{teo:classbetas}
When $0\le \beta<1/2$
 \begin{equation}
 \label{classbetas}
 \lim_{N\to\infty\vert \beta}
 \sum_{C_{N,\overline {M}}}
 P^{{\cal{R}}-{\cal{P}}}_{N,\overline M,\overline L}(C_{N,\overline {M}})
 \frac{T_{\overline {M},C_{N,\overline {M}},p}}{N p |\log p|}= \frac{1}{{(\log 2)}^2}.
 \end{equation}
 \end{theorem}

\vspace{0.2 cm}

Finally, we provide
a random class of connectivity matrices
for which our upper bound in (\ref{asymptoticblarge}) for the case
$p=1/N^{\beta}$ with $1/2\leq\beta<1$ is attained. Let
$P^{{{\cal{P}}-{\cal{P}}}}_{N,M,L}$ denote the distribution
of random bipartite graphs with $N$ variable nodes, $M$ test nodes
and $l$ tests  per variable ($k$ variables per test), with $l$ 
Poisson distributed and with mean $L$ ($k$ Poisson distributed with mean $K=NL/M$), namely
\begin{equation}
\label{poipoi}
P^{{{\cal{P}}-{\cal{P}}}}_{N,M,L}(C_{N,M})=
\prod_{a=1}^M\prod_{i=1}^N\left(\frac{L}{M}\right)^{c_{i,a}}\left(1-\frac{L}{M}\right)^{1-c_{i,a}}.
\end{equation}
If we make the choice $L=\widetilde L$ and $M=\widetilde M$ with
\begin{equation}
\label{tildel}
\widetilde L=:[e|\log p|]
\end{equation}
\begin{equation}
\label{tildem}
\widetilde M=:[e Np|\log p|]
\end{equation}
the following result, whose proof is given in section \ref{upper2}, holds

\vspace{0.2 cm}

\begin{theorem}
\label{teo:classbetal}
When $0\leq\beta<1$
\begin{equation}
\label{bsmall}
\lim_{N\to\infty\vert \beta} 
\sum_{C_{N,\widetilde{M}}}
P^{{\cal{P}}-{\cal{P}}}_{N,\widetilde M,\widetilde L}(C_{N,\widetilde {M}})
\frac{T_{\widetilde{M},C_{N,\widetilde {M}},p}}{N p |\log p|}= e.
\end{equation}
\end{theorem}

\vspace{0.2 cm}

Furthermore, 
the choice (\ref{tildel})-(\ref{tildem}) for the couple $(M,L)$ is optimal over all the Poisson-Poisson distributions, namely
\begin{remark}
\label{byside}
When $0\leq\beta<1$, for any $M,L$ 
\begin{equation}
\lim_{N\to\infty\vert \beta} 
\sum_{C_{N,\overline {M}}}
P^{{\cal{P}}-{\cal{P}}}_{N,M,L}(C_{N,M})
\frac{T_{M,C_{N,{M}},p}}{N p |\log p|}\geq e.
\end{equation}
\end{remark}

Note that the Poisson-Poisson 
distribution  had already been used in \cite{Berger} to obtain the
upper bound  on $\overline T(N,p)$ which we have recalled in formula (\ref{bergerbound}). 
Here, 
by optimizing 
the choice of the parameters $L$ and $M$ for $P^{{\cal{P}}-{\cal{P}}}_{N,M,L}$,
we ameliorate the upper bound (\ref{bergerbound}) which was obtained in  \cite{Berger}
by choosing $M=4Np|\log(p)|$ and $L=2|\log p|$.

Furthermore the result of Remark \ref{byside} together with
Theorem \ref{teo:betas}
imply that the optimal value for $\beta\in[0,1/2)$ can never be attained on the class of Poisson-Poisson distributions, while it can be attained both on the class of regular-Poisson distributions and on the class of regular-regular graphs with girth at least $6$ (see Theorem  \ref{teo:classbetas}).

\vspace{0.5 cm} The outline of the paper is the following. 

In section \ref{lower} we
establish a lower bound on $T_{M,C_{N,M},p}$ (Theorem \ref{th:lower}) which
holds for any $N$ and $p\to 0$. In section \ref{regreg} we prove Theorem
\ref{teo:regbetas} which, together with Theorem \ref{th:lower}, completes the
proof of Theorem \ref{teo:betas} and identifies a set of algorithms for which
the asymptotic value of $T_{M,C_{N,M},p}$ is attained if $\beta\in[0,1/2)$. In
section \ref{upper} we prove Theorem \ref{teo:classbetas} which identifies a
different class of random algorithms over which this asymptotic value is also
attained. In Section \ref{upper2} we prove Theorem \ref{teo:classbetal} which,
together with Theorem \ref{th:lower}, completes the proof of Theorem
\ref{teo:betal} and identifies an algorithm over which our upper bound on
$T_{M,C_{N,M},p}$ is attained when $1/2\leq\beta<1$.

\section{Lower bound on $\overline{T}(N,p)$}
\label{lower}

In this section we prove the following 
lower bound on $\overline T$ which 
holds for any $N$ whenever we let
$p\to 0$.

\vspace{0.2 cm}

\begin{theorem}
\label{th:lower}
\begin{equation}
\label{eqnlower}
\lim_{p\to 0}
\frac{\overline{T}(N,p)}{{Np|\log p|}}\geq \frac{1}{(\log 2)^2}.
\end{equation}
\end{theorem}

\vspace{0.2 cm}
When one takes the limit $\lim_{N \to \infty\vert \beta}$ with
 $\beta\in [0,1)$,
 Theorem \ref{th:lower}
improves the previously existing lower bounds \cite{Berger} on $\overline{T}(N,p)$.
Furthermore for all the cases $\beta\in[0,1/2)$ it allows, together with Theorem \ref{teo:classbetas},
to determine the exact value of 
$\lim_{N\to\infty\vert \beta}\overline T/(Np|\log p|)$.
On the other hand when $\beta\geq 1$
better bounds then the one given by our (\ref{eqnlower}) 
already exist \cite{Knill,Berger}.

Let $q:=(1-p)$ and $Z^+_{N}:=\{0,1,\dots,N\}$, we define 
$B_{M,C_{N,M},p}$ and the function $A_p:(Z^+_N)^N\to\mathbb R$ as
\begin{IEEEeqnarray}{l}
\label{defB}
B_{M,C_{N,M},p}:= 
\displaystyle{q\sum_{i=1}^N\prod_{a=1}^M \left(1-q^{d_a-1}\right)^{c_{i,a}}}
\\
\label{original}
A_p(\vec m):=\sum_{i=1}^N \frac{m_i}{i}+q 0^{m_1}\exp{(-\sum_{i=2}^N m_i a^i_p)}\nonumber\\
\end{IEEEeqnarray}
where $d_a$ is the degree of test $a$, i.e.
\begin{equation}
\label{da}
d_a:=\sum_{l=1}^N  c_{l,a}
\end{equation}
and, for $i=2,\dots N$,
$$a^i_p:=|\log\left(1-(1-p)^{i-1}\right)|.$$ 
We also let
\begin{IEEEeqnarray}{l}
\label{minA}
\bar A_p:=\min_{\vec m\in(Z^+_N)^N}A_p(\vec m)\\
U(p):=\min_{r\in [2,\infty)} \; \frac{1}{r \vert \log(1-(1-p)^{r-1}) \vert}
\label{Udef}\\
c(p):=\min_{w\in [0,\infty)} \; \left(U(p) w+(1-p)e^{-w}\right)
\label{ancora}
\end{IEEEeqnarray}

In order to prove Theorem \ref{th:lower} we will use the following 
results, whose proofs are postponed to the next sections.

\vspace{0.2 cm}

\begin{Lemma}
\label{lemma2}
$$~$$

\vspace{-0.3 cm}
(a) For any choice of $M$ and $C_{N,M}$ the expected number of undetected zeros, ${\cal{U}}_0$, is lower bounded by

\begin{equation}
\label{FKGsure}
\displaystyle{\sum_{X}\mu_p(X)|{\cal{U}}_0|}\geq B_{M,C_{N,M},p}.
\end{equation}

(b) If the girth of the graph ${\mathcal G}(C_{N,M})$ 
is larger or equal to $6$, (\ref{FKGsure}) holds as an equality.
\end{Lemma}

\vspace{0.2 cm}

\begin{Lemma}
\label{lemma3}
$$\min_{M,C_{N,M}} (M+B_{M,C_{N,M},p})\geq  N \bar A_p.$$

\end{Lemma}

\vspace{0.2 cm}

\begin{Lemma}
\label{lemma4}
\begin{equation}
\label{minAbis}
\bar A_p\geq  \min(1,c(p)).
\end{equation}
\end{Lemma}

\vspace{0.2 cm}

\begin{claim}
\label{claim}
When $p\to 0$, $U(p)=\frac{p}{(\log 2)^p}+\Theta(p^2)$. 
\end{claim}

\vspace{0.4 cm}

\IEEEproof [Proof of Theorem \ref{th:lower}]

By using
 definition
(\ref{meantest}), Lemma  \ref{lemma2} and the trivial inequality
$\sum_X\mu_p(X)|{\cal U}_1|\geq 0$,
we  get 
\begin{equation}
\label{tineq}
T_{M,C_{N,M},p}\geq M+B_{M,C_{N,M},p}.
\end{equation}
Notice that for all the choices of $p$ that we consider,
the bound will not suffer from the fact 
that we neglected the contribution from 
undetermined ones. This can be seen from the facts
$$\sum_{X}\mu_p(X)|{\cal{U}}_1|\leq \sum_{X}\mu_p(X) |{\cal{D}}|=pN$$ 
and 
$T_{M,C_{N,M},p}\geq -Np\log_2 p$ (this is the information
theoretic lower bound \cite{Berger}),which imply that 
for any $\beta \in [0,1)$:
$$\lim_{N\to\infty \vert \beta}\frac{\sum_{X}\mu_p(X)|{\cal{U}}_1|}{T_{M,C_{N,M},p}}=0.$$

Since (\ref{tineq}) holds for any $(M,C_{N,M})$, by using Lemma \ref{lemma3} and \ref{lemma4} it follows immediately that
\begin{equation}
\label{4un}
\overline{T}(N,p)\geq 
N\min(1,c(p)).
\end{equation}

From definition (\ref{ancora}), as an immediate corollary of Claim \ref{claim},
we get 
\begin{equation}
\label{c1}
c(p)=\frac{p|\log p|}{(\log 2)^2}+(1-2|\log(\log 2)|)\frac{p}{(\log 2)^2}+o(p)
\end{equation}
 in the limit $p\to 0$.
By gathering the results (\ref{4un}) and (\ref{c1})
the proof of Lemma (\ref{lower}) is concluded.
 Furthermore, we get
the following lower bound for the corrections
\begin{equation}
\label{lowercorr}
\frac{1-2|\log(\log 2)|}{(\log 2)^2}
\leq \frac{\overline T(N,p)-Np|\log p|(\log 2)^{-2}}{Np}
\end{equation}
\endproof

\subsection{Proof of Lemma \ref{lemma2}}

(a) By definition the set of undetected zeros, ${\cal{U}}_0$, contains 
all the variables $i$ such that $x_i=0$ and $T_a=1$ 
for any $a$ such that $c_{i,a} =1$, i.e. $i$ belongs only to pools containing at least a variable equal to one.
Therefore:
\begin{equation}
\label{aiha}
\displaystyle{\sum_{X}\mu_p(X)|{\cal{U}}_0|}=
\displaystyle{\sum_{i=1}^N\sum_X\mu_p(X)(1-x_i)\prod_{a=1}^M W_{i,a}(X)}
\end{equation}
where 
\begin{equation}
\label{wia}
W_{i,a}(X):=
\left(1-\prod_{\stackrel{j=1,\dots N}{j\neq i}}(1-x_j)^{c_{j,a}}\right)^{c_{i,a}}.
\end{equation}
Since $W_{i,a}(X)$ does not depend on $x_i$  we can immediately perform the average over this variable for each term of the sum in (\ref{aiha}).
Then, for each given $i$, we introduce the partial  order $\prec_i$ according to which
$x\prec_i x'$ if and only if $x_j\leq x'_j$ for all $j\in\{(1,\dots,N)\setminus i\}$.
For any $C_{N,M}$ and for
any $a\in(1,\dots, M)$,
$W_{i,a}$ is a  non-decreasing function
with respect to this partial order, namely 
$x\prec_i x'$ implies that $W_{i,a}(X)\leq W_{i,a}(X')$.
Therefore inequality (\ref{FKGsure}) follows 
by applying FKG inequality \cite{FKG} to each term of the sum  in 
(\ref{aiha}). In other words, we have simply used
the positive correlation among the events that there exists at least one 
variable equal to one in two (or more) intersecting 
pools.

(b) If the
bipartite graph has girth at least $6$, i.e. if 
$M,C_{N,M}$ are such that 
for any couple of variables there exists at most one test
which contains both of them (see condition (\ref{conditionmatrix})),
the events defined above are independent. Therefore 
(\ref{FKGsure}) holds with the equality sign.
\endproof

\subsection{Proof of Lemma \ref{lemma3}}

Given a choice $(M,C_{N,M})$,  we define for each variable $i$ 
the vector 
$\vec {m}^i=(m^i_1,\dots,m^i_N)\in (Z^+_N)^N$
where $m^i_j$ denotes the number of tests which contain variable $i$ and 
globally contain $j$ variables 
$$m^i_j:=\sum_{a=1}^M c_{i,a} \delta_{j,d_a}$$
where $d_a$ is defined in \eqref{da}.
(and $\delta$ is the
Kronecker $\delta$).
Then we define for each  
$\vec m=(m_1,\dots,m_N)\in (Z^+_N)^N$ 
a density, $f(\vec m)$, such that $N f(\vec m)$ is the number of variables 
$i$ for which $\vec m^i=\vec m$. 
With this notation we can rewrite the number of tests, $M$, 
 and the definition
 (\ref{defB}) for $B_{M,C_{N,M},p}$  as
\begin{IEEEeqnarray}{l}
\label{Mrew}
M=N\sum_{\vec m}f(\vec m)\sum_{j=1}^N \frac{m_j}{j}\\
\label{Brew}
B_{M,C_{N,M},p}=N q \sum_{\vec m} f(\vec m) P(\vec m) 
\end{IEEEeqnarray}
where here (and whenever it appears in the following) 
the sum over $\vec m$ is performed on $\vec m\in (Z^+_N)^N$
and
\begin{equation}
P(\vec m):=
\displaystyle{
\prod_{j=1}^N (1-q^{j-1})^{m_j}}.
\end{equation}
Let also
$$e_j:=\sum_{a=1}^M c_{i,a} \delta_{j,d_a}$$
with this notation and using \eqref{Mrew} and \eqref{Brew} it is immediate to check that
\begin{equation}
\label{4bis}
M+B_{M,C_{N,M},p}=\tilde B_p(f_{M,C_{N,M}})
\end{equation}
where, for any couple $(M,C_{N,M})$,
$f_{M,C_{N,M}}:(Z^+_N)^N\to(0,1/N,\dots N/N)$ is defined by
\begin{IEEEeqnarray}{l}
\label{cacio}
f_{M,C_{N,M}}(\vec m):=
N^{-1}\sum_{i=1}^N \prod_{j=1}^N
\delta_{m_j,e_j}
\end{IEEEeqnarray}
and
\begin{IEEEeqnarray}{l}
\label{newdefB}
\displaystyle{\tilde B_p(f):=}\\ 
\displaystyle{N\sum_{\vec m}f(\vec m)\left[\sum_{i=1}^N \frac{m_i}{i}+
q 0^{m_1}\exp(\!{-\sum_{i=2}^N m_i a^i_p})\right]}\nonumber
\end{IEEEeqnarray}
Therefore by using (\ref{4bis}) and definition (\ref{original})
we get
\begin{IEEEeqnarray}{c}
\label{tineq2}
\min_{M,C_{N,M}} (M+B_{M,C_{N,M},p})=\min_{M,C_{N,M}}\tilde B_p(f_{M,C_{N,M}})\nonumber\\
\geq \inf_{f\in{\mathcal F}}\tilde B_p(f)\geq N\min_{\vec m\in (Z^+_N)^N}A_p(\vec m)=N \bar A_p\nonumber\\
\end{IEEEeqnarray}
where 
 ${\mathcal F}$ is the set of probability functions 
on $(Z^+_N)^N$, i.e. $f:(Z^+_N)^N\rightarrow{\mathbb R}^+$ with $\sum_{\vec m}f(\vec m)=1$.
The second inequality immediately follows from the definition (\ref{original}) and the fact that $f$ is a probability distribution.
\endproof

The following remark will be used to construct 
 an optimal algorithm in section \ref{regreg}

\begin{remark}
\label{remarkonmin}
Define $g\in\mathcal F$ as

\begin{numcases}
{g(\vec m):=}
\nonumber
1  & {\mbox{if}} $\vec m=\overline m$\\
0 & {\mbox{otherwise}}
\end{numcases}
with $\overline m\in (\mathbb Z^+)^N$ such that

\begin{numcases}
{\overline m_i:=}
\nonumber
[{|\log p|}/{\log 2}]  & {\mbox{if}} $i=[\log 2/p]$\\
0 & {\mbox{otherwise}}
\end{numcases}

Then:
\begin{equation}
\frac{\tilde B_{p}(g)}{Np|\log p|}=\frac{1}{(\log 2)^2}+o(p).
\end{equation}
Furthermore $g$
coincides with $f_{M,C_{N,M}}$ (\ref{cacio})
on all bipartite graphs with
$M=N[p|\log p|/(\log 2)^2]=\overline M$
tests and  connectivity matrix $C_{N,M}$ such that the 
 number of tests per variable is fixed equal to $[{|\log p|}/{\log 2}]=\overline L$.

\end{remark}

\subsection{Proof of Lemma \ref{lemma4}  and Claim \ref{claim}}

\IEEEproof [Proof of Lemma \ref{lemma4}]
Let $\overline m$ be the vector
over which $\bar A_p$ is reached.
 We consider separately the two complementary cases:
(a)  $\overline m_1\geq 1 $
and (b)  $\overline m_1= 0$.
In case (a), the minimum is obviously larger or equal to one. Therefore
\begin{equation}
\label{minAbis}
\bar A_p\geq \min(1,b_p)
\end{equation}
where 
$$b_p:=\min_{\stackrel{\vec m\in(Z_{N})^N}{ m_1=0}}A_p(\vec m)$$

We now enlarge  the  minimization of $m_i$
to all real positive values $\mathbb R^+=[0,\infty)$, and introduce the
two functions on $ (\mathbb R^+)^{N-1}$:
\begin{eqnarray}
u(\vec m)&:=&\sum_{i=2}^N m_i/i \\
v_p(\vec m)&:=&\sum_{i=2}^N m_i a^i_p
\end{eqnarray}
A simple bound on $b$ is expressed in terms of these functions:
\begin{equation}
b_p\geq
\min_{\vec m\in (\mathbb R^+)^{N-1}} \left(
u(\vec m) +(1-p) e^{-v_p(\vec m)}\right). 
\label{buv}
\end{equation}
This minimization is carried out in two steps. We first fix $v_p(\vec m)=w\geq 0$, and look
for the minimum of $u$ in the subspace $v_p(\vec m)=w$. Let us denote by $u^*(w,p)$ this minimum value.
 Finding $u^*$ is a problem of linear optimization. So the minimum must be obtained on one of the vertices of the simplex of $ (\mathbb R^+)^{N-1}$ defined by $v_p(\vec m)=w$. These vertices are easily identified:
There are $N-1$ of them, located at points $\vec m^{(2)},\dots , \vec m^{(N)}$, with
$ m^{(r)}_j=\delta_{j,r} w/a^r_p$. As $u(\vec m^{(r)})=w/(r a^r_p)$ , the minimum of $u$ is at $u^*(w,p)=w\min_{r\in\left\{2,\dots,N\right\}}1/(r a^r_p)$.
By enlarging the space of $r$ to all real values in $[2,\infty)$, we get:
\begin{equation}
u^*(w,p)\geq w U(p).
\label{ulam}
\end{equation}
Now we carry the optimization in (\ref{buv}) as:
\begin{equation}
b_p\geq
\min_{w\in[0,\infty)}\left[w U(p)+(1-p) e^{-w}\right]
\label{bw}
\end{equation}
which establishes lemma \ref{lemma4}.
\endproof

\vspace{0.2 cm}

\IEEEproof [Proof of Claim \ref{claim}]
Let $z:=(1-p)^{r-1}$ and
\begin{equation}
g_p(z):=\frac{(1-z)\log(1-z)}{z\log[z(1-p)]}.
\end{equation}
For $r\geq 2$ and $0<p<1$,  it is immediate to verify that $0<z\leq (1-p)$ and that
any stationarity point for $U(p)$ must satisfy $g_p(z)=1$.
By studying the function $g_p(z)$ it is then possible 
to prove that there are 
two values $z\in (0,1)$ which satisfy the latter condition. Furthermore, when
$p\to 0$ only one of these two values belongs to $(0,1-p]$
and it corresponds to $z=1/2-\epsilon(p)$
 with $\epsilon(p)=\Theta(p)$. The desired result for $U(p)$ immediately follows.
\endproof

\section{Upper bounds on $\overline{T}(N,p)$ for $\beta\in[0,1/2)$ via regular-regular graphs}
\label{regreg}

In this section we prove Theorem \ref{teo:regbetas}. On
the one hand this result allows
to complete the proof of Theorem
\ref{teo:betas}, namely to identify the sharp asymptotic value of
$\overline T(N,p)/(Np|\log p|)$ in the limit $N\to\infty$, $p=N^{-\beta}$ for
$\beta\in[0,1/2)$. Precisely,

\vspace{0.2 cm}

\IEEEproof [Proof of Theorem \ref{teo:betas}]
The proof follows immediately from Theorem \ref{th:lower} and Theorem 
 \ref{teo:regbetas}.
\endproof

\vspace{0.2 cm}

On the other hand Theorem  \ref{teo:regbetas}
provides a constructive procedure for
 a class of algorithms (i.e. a choice of $M$ and a class of matrices
$\{C_{N,M}\}$) which are asymptotically optimal.

In order to  construct these algorithms we will keep in mind the following observations. 
First, as already remarked in the  previous section, 
the number of tests due to the undetermined ones is negligible 
for all the choices of $p$ discussed in Theorem \ref{teo:regbetas}.
Therefore we focus on 
algorithms that minimize the number of tests in the first stage, $M$, plus the number of undetermined zeros, $|{\cal{U}}_0|$. 
The second observation is that 
inequality (\ref{tineq}) comes from 
({\ref{FKGsure}) and the latter becomes
an equality provided $M,C_{N,M}$ are such that the corresponding graph has girth at least $6$.
The third observation is the one contained in remark \ref{remarkonmin} which states that
the minimum for the right hand side of (\ref{tineq}) is  attained
on any graph with $\overline M$ tests and $\overline L$ tests per variable, where $\overline M$ and $\overline L$ have been defined in
(\ref{l}) and (\ref{m})  (we recall that both $\overline M/N$ and $\overline L$ depend only on $p$). 
Therefore {\sl if} it is possible to find at least one graph with 
 girth at least $6$ among those with $\overline M$ tests and
$\overline L$ tests per variable, the mean number of tests on this graph will match the lower bound in Theorem \ref{th:lower}  
in the limit $p\to 0$. 
In the following we will use the above ideas and
the results on regular-regular graphs with a fixed minimal girth
which have been obtained in \cite{LuMoura}.

\begin{IEEEproof}[Proof of Theorem \ref{teo:regbetas}]
 Consider a connectivity matrix $C_{N,\overline M}^{\overline L}$ with fixed
  variable degree $\overline L$, fixed test degree $\overline K=\overline
  LN/\overline M$ and girth at least $6$ (see condition
  \eqref{girthcondition}). The proof of the existence of such a graph and an
  explicit procedure for its construction have been provided by Lu and Moura in
  their study of large girth LDPC codes \cite{LuMoura}.
 Their procedure requires that  the condition 
\begin{equation}
\label{girthcondition}
\overline M\geq\frac{(\overline L-1)(N\overline L/\overline M)}{\overline L\overline K-\overline L-\overline K}
\end{equation}
be satisfied. (This condition corresponds to condition (14) in
appendix A of \cite{LuMoura} for the choice  $g=6$). In the limit $N\to\infty$ with $p=1/N^{\beta}$ with $\beta<1/2$, 
the validity of 
(\ref{girthcondition}) can be readily checked, using definitions \eqref{l} and \eqref{m}.

From equation \eqref{meantest} and \eqref{aiha}, 
the number of tests on any such graph
satisfies the inequality
\begin{equation}
\label{upperreg}
\lim_{N\to\infty}
T_{\overline {M},C_{N,\overline {M}}^{\overline L},p}
\leq
\overline M+Np+N(1-p)R_p
\end{equation}
where
\begin{IEEEeqnarray}{l}
\label{Rp}
R_p=\sum_{X}\mu_p(X)\prod_{a=1}^{\overline M}W_{i,a}
=(1-(1-p)^{\overline K-1})^{\overline L}.\nonumber\\
\end{IEEEeqnarray}
The last equality is obtained by using definition (\ref{wia}) for $W_{i,a}$
(the mean over the Bernoulli distribution
is  easily performed thanks to the girth condition).
Theorem  \ref{teo:regbetas}
immediately follows from \eqref{upperreg}.
\end{IEEEproof}

\vspace{0.2 cm}

The results \eqref{upperreg} and \eqref{Rp}
hold for any choice of p. However, it is important to notice
that the existence of at least one such  connectivity 
matrix with girth at least $6$ is guaranteed only for $\beta\in[0,1/2)$. 
In particular for $p=N^{-\beta}$ with $1<\beta<2$
 there cannot exist any such  matrix: otherwise \eqref{upperreg} and \eqref{Rp} would imply
$\overline T(N,p)\leq \beta/(\log 2)^2 N^{1-\beta}\log N$ 
which goes to zero as $N\to\infty$ (and
is in contradiction with the lower bounds
in  formula (57) of \cite{Berger}).

Putting together \eqref{upperreg}, \eqref{Rp} and \eqref{lowercorr}
it is also immediate to verify that the higher order corrections to the optimal value $\overline T(N,p)$ are of  order $pN$. More precisely, if we let
\begin{equation}
\label{defcorrez}
H(N,p):=\frac{\overline T(N,p)-Np|\log p|(\log 2)^{-2}}{Np}
\end{equation}

the following holds.

\vspace{0.2 cm}

\begin{remark}
\label{corrections}
For $\beta\in[0,1/2)$ in the limit $N\to \infty$ the following holds

\begin{equation}
\frac{1-2|\log\log 2|}{(\log 2)^2}
\leq H(N,p)\leq 2
\end{equation}
\end{remark}

\section{Regular-Poisson graphs 
are (also) optimal for $\beta\in[0,1/2)$}
\label{upper}
 In this section we will
 prove Theorem \ref{teo:classbetas} which shows that
 asymptotically optimal pool designs are obtained with regular-Poisson
 distributions for proper choices of the graph parameters. This is
 particularly relevant since the construction of  regular-Poisson graphs
is much simpler than the  construction
 of \cite{LuMoura} for regular-regular graphs with girth at least $6$.

Consider the  {\sl regular-Poisson} distribution on bipartite graphs  defined in section \ref{results} with $N$ variable nodes, $\overline M$ test nodes and $\overline L$ tests per variable, $P ^{{\cal {R}}-{\cal{P}}}_{N,\overline M,\overline L}$.
Fix a variable, $i$, and let ${\cal{E}}_i^n$ be the characteristic function of the event 
(defined over the space of all bipartite graphs with $\overline M$ nodes) that there are more than $n$ loops of length $4$ which contain $i$,
i.e. there are more than $n$ triples $(j,a,b)$ with $j$ a variable different from $i$ ($j\neq i$, $j\in(1,\dots,N)$) and $a,b$ two distinct tests ($a\neq b$, $a,b\in (1,\dots \overline M)$) such that $i$ and $j$ belong to both tests.
 Precisely, we define
 ${\cal{E}}_{i}^n:C_{N,\overline M}\in\{0,1\}^{N\times \overline M}\to\mathbb R$ as
${{\cal{E}}_{i}^n  (C_{N,\overline M}):=}1$ if
\begin{equation}
\displaystyle \sum_{j=1,(j\neq i)}^N\sum_{1\le a<b\le M}\;  c_{i,a}c_{i,b}c_{j,b}c_{j,a}> n
\end{equation}
 and  ${{\cal{E}}_{i}^n  (C_{N,\overline M}):=}0$ otherwise.

In order to prove Theorem \ref{teo:classbetas} we will need the following Lemmas  which give an upper bound on the probability that there are more than $n$ loops of length $4$ through $i$ (Lemma \ref{lemma7})
and an upper bound on the probability that $i$ is an undetermined zero and
does not belong to more than $n$ loops of length $4$ (Lemma \ref{lemma9}).
\vspace{0.2 cm}

\begin{Lemma} 
\label{lemma7}
\begin{eqnarray}
\label{loops}
&&\sum_{C_{N,\overline M}} P^{{\cal {R}}-{\cal{P}}}_{N,\overline M,\overline L}(C_{N,\overline M})
{{\cal{E}}_{i}^n}(C_{N,\overline M})
\leq\nonumber\\
&&\leq \frac{N\overline L^6}{\overline M^3}+
 \left(\frac{N\overline L^4}{\overline M^2}\right)^{n+1}.
\end{eqnarray}

\end{Lemma}

\vspace{0.2 cm}

\vspace{0.2 cm}

\begin{Lemma}
\label{lemma9}
Let 
$\overline k$ be the average degree of the checks,
$\overline k:= N \overline L/\overline M=\log 2/ p$,
and
\begin{equation}
C_p:= \sum_{C_{N,\overline M}}\!P^{{\cal{R}}-{\cal{P}}}_{N,\overline M,\overline L}(C_{N,\overline {M}})(1-{{\cal{E}}_{i}^n})\sum_{X}\mu_p(X)\prod_{a=1}^{\overline M} W_{i,a}
\label{Cp}
\end{equation}
(we drop for semplicity of notation the dependence
of $W_{i,a}$ on $X,C_{N,\overline M}$ and the dependence
of ${\cal{E}}_{i}^n$ on $C_{N,\overline M}$).
Define also $\gamma:=p^{\alpha}$.

For any $n$ and $\alpha$ with $0<n<\overline L/2$ and $0<\alpha<1/2$, 
the following holds 
\begin{IEEEeqnarray}{l}
C_p\leq 
\left[1-(1-p)^{\overline k (1+\gamma)}\right]^{\overline L-2n}+\\
+\overline L\exp\left[-\frac{\gamma^2\log 2}{2p}\right]+o\left(\overline L
\exp\left[-\frac{\gamma^2\log 2}{2p}\right]\right).\nonumber
\label{now}
\end{IEEEeqnarray}

\end{Lemma}

\vspace{0.2 cm}

\vspace{0.2 cm}

 \begin{IEEEproof}[Proof of Theorem  \ref{teo:classbetas}]

For any $n$ and $\alpha$ with $0<n<\overline L/2$
and $0<\alpha<1/2$, the  mean number of tests verifies
\begin{IEEEeqnarray}{l}
\label{upperf}
\sum_{C_{N,\overline {M}}}
 P^{{\cal{R}}-{\cal{P}}}_{N,\overline M,\overline L}(C_{N,\overline {M}})T_{\overline {M},C_{N,\overline {M}},p}
\leq \overline M+Np+ \nonumber\\
\frac{|\log p|^3}{Np^3}+N\left(\frac{|\log p|^2}{Np^2}\right)^{n+1}\!\!+
N\left[1-(1-p)^{\overline k (1+\gamma)}\right]^{\overline L-2n}\nonumber\\
+N\overline L\exp\left[-\frac{\gamma\log 2}{2p}\right]+No\left(\overline L
\exp\left[-\frac{\gamma\log 2}{2p}\right]\right)\nonumber\\
\end{IEEEeqnarray}
where $\gamma:=p^{2\alpha}$.

In order to derive \eqref{upperf} we have: (i) used definition (\ref{meantest});
(ii) bounded the mean number of undetermined ones with the mean number of ones;
(iii) decomposed the mean number of undetermined zeros into those that are
(are not) on a variable node $i$ which contain more than $n$ loops of length 4; (iv) upper
bounded the last two terms via the results of Lemma \ref{lemma7} and Lemma \ref{lemma9}.
If we now make the choice $n:=\overline L^{-1/2}$ the result
of the Theorem immediately follows by noticing that:
\begin{IEEEeqnarray}{l}
\lim_{N\to\infty\vert \beta}\!\!\frac{N\overline L\exp(-\gamma^2p^{-1}\log 2/2)}{Np|\log p|}=0\\
\lim_{N\to\infty\vert \beta}\frac{|\log p|^3}{Np^3 N p|\log p|}=0\\
\lim_{N\to\infty\vert \beta}
\frac{N}{Np\log p}\left(\frac{|\log p|^2}{Np^2}\right)^{n+1}=0.
\end{IEEEeqnarray}
%note that last limit is 0 for n> A=3beta-1/1-2beta but A<(1-2beta)^{-1}
%thus n>A when Nto infty since n sqrt L
%CR
%which follows immediately from our choice
%our choice $\gamma=p^{\alpha}$ with $\alpha<1/2$.
%
\end{IEEEproof}

\vspace{0.2 cm}

\begin{IEEEproof}[Proof of Lemma \ref{lemma7}]
Given a connectivity matrix, $C_{N,\overline M}$, 
we identify among the loops of length $4$ 
two distinct classes: loops of type
$S$ and of type $D$. Loops of type $S$  are 
those disconnected from
any other loop, namely they correspond to the
 choices of two variables $i,j$ and two tests $a,b$
such that both $i$ and $j$ belongs to $a$ and $b$ and there does not exist another test containing both $i$ and $j$. Loops of type $D$ are all loops of length $4$ which are not of type $S$.
For a given variable $i$, let ${\cal{D}}_i$ be the characteristic function of the event 
that $i$ belongs to at least one loop of type $D$.
 Precisely, we define
 ${\cal{D}}_{i}:C_{N,\overline M}\in\{0,1\}^{N\times \overline M}\to\mathbb R$ as
${{\cal{D}}_{i}  (C_{N,\overline M}):=}1$ if
\begin{equation}
\displaystyle \sum_{j=1,(j\neq i)}^N\sum_{1\le a<b<c\le M}\;  c_{i,a}c_{i,b}c_{i,c}c_{j,b}c_{j,a}c_{j,c}>0
\end{equation}
 and  ${{\cal{D}}_{i}(C_{N,\overline M}):=}0$ otherwise. 

The following inequalities hold
\begin{IEEEeqnarray}{l}
\label{loopsD}
\sum_{C_{N,\overline M}} P^{{\cal {R}}-{\cal{P}}}_{N,\overline M,\overline L}(C_{N,\overline M})
{{\cal{D}}_{i}}\leq
\frac{N\overline L^6}{\overline M^3}\\
\label{loopsS}
\sum_{C_{N,\overline M}} P^{{\cal {R}}-{\cal{P}}}_{N,\overline M,\overline L}(C_{N,\overline M}){{\cal{E}}_{i}^n}
\left(1-{{\cal{D}}_{i}}
\right)
\leq \left(\frac{\overline L^4 N}{\overline M^2}\right)^{n+1}
\end{IEEEeqnarray}
(we drop the dependence of ${{\cal{D}}_{i}}$ and ${{\cal{E}}_{i}}^n$ on $C_{N,\overline M}$.)
%is derived by an upper bound on the number of possible D loops which contain $i$ times the upper bound on the probability that  one such loops is verified.
%is derived by an upper bound on the number of possible $n+1$-ples of 
%S loops which contain $i$ times an upper bound on the probability that all such looops are present (which uses the fact that 
%any of tese two loops do not have in common the variable $j\neq i$) 
The result follows immediately from \eqref{loopsD} and \eqref{loopsS} and the fact that, for any $i$ and $C_{N,\overline M}$, ${{\cal{E}}_{i}}^n(C_{N,\overline M})<1$.
\end{IEEEproof}

\begin{IEEEproof}[Proof of Lemma \ref{lemma9}]
For a given connectivity matrix $C_{N,\overline {M}}$, let $i$ be a site with less than $n$ loops
of length $4$. Let us call $A_i({C_{N,\overline {M}}})$ the set of tests which contain $i$,
$B_i({C_{N,\overline {M}}})$ the set of tests which belong to a loop of length  $4$ passing through $i$, and
 $\overline B_i({C_{N,\overline {M}}})=A_i({C_{N,\overline {M}}})\setminus B_i({C_{N,\overline {M}}})$. 
Clearly $\vert B_i({C_{N,\overline {M}}})\vert\le 2n$.
The following holds
\begin{IEEEeqnarray}{l}
\label{anco}
\prod_{a=1}^{\overline M} W_{i,a}\le \prod_{a\in \overline B_i({C_{N,\overline {M}}})} W_{i,a} \\
= \prod_{a\in \overline B_i(C_{N,\overline {M}})}
\left(1-\prod_{\stackrel{j=1,\dots N}{j\neq i}}(1-x_j)^{c_{ja}}\right).\nonumber\\
\end{IEEEeqnarray}
We can now plug
\eqref{anco} into \eqref{Cp} and  get
\begin{IEEEeqnarray}{l}
C_p\le \sum_{C_{N,\overline M}}\!P^{{\cal{R}}-{\cal{P}}}_{N,\overline M,\overline L}(C_{N,\overline {M}})\\
 \id (|\overline B_i({C_{N,\overline {M}}})|\ge \overline L-2n)\prod_{a \in  \overline B_i(C_{N,\overline M})}\! [1-(1-p)^{d_a-1}]
\label{newC}
\nonumber
\end{IEEEeqnarray}
where in order to perform the mean over $\mu(X)$ we used the fact that 
the neighborhoods of any two tests $a,b$ belonging to $\overline  B_i(C_{N,\overline {M}})$ 
 intersect only in $i$
(for any $j \ne i$ one has $c_{ja}c_{jb}=0 $) and we recall from definition \eqref{da} that $d_a$ is the degree of test $a$. 
%CR Not so nice/precise this P_{\rm max}, I erase and change below
%Let us introduce the probability $P_{\rm max}$ which is the probability (with respect to $P^{{\cal{R}}-{\cal{P}}}_{N,\overline M,\overline L}(C_{N,\overline {M}})$)
% of the maximum degree of the first $L$ tests,
% namely $k_{max}:=\max_{a\in(1,\dots,\overline L)}\sum_{j} c_{j,a}$.
%Using the invariance of the regular-Poisson distribution under test permutations, we get:
%\begin{IEEEeqnarray}{l}
%\label{Cp3}
%C_p\leq 
%\sum_{k=0}^{N} P_{\rm max}(k)(1-(1-p)^{k})^{\overline L-2n}\nonumber\\
%\leq (1-(1-p)^{\bar k(1+\gamma)})^{\overline L-2n}+\overline L D_p nonumber\\
%\end{IEEEeqnarray}
Let $k_{max}$ be the maximum degree of the first $\overline L$ tests,
namely $k_{max}:=\max_{a\in(1,\dots,\overline L)}\sum_{j} c_{j,a}$.
Using \eqref{newC} and the invariance of the regular-Poisson distribution under test permutations, we get
\begin{IEEEeqnarray}{ll}
\label{Cp3}
C_p&\leq \sum_{C_{N,\overline M}}\! P^{{\cal{R}}-{\cal{P}}}_{N,\overline M,\overline L}(C_{N,\overline {M}})\sum_{k=0}^{N} \delta_{k,k_{\rm max}}(1-(1-p)^{k})^{\overline L-2n}\nonumber\\
&\leq (1-(1-p)^{\bar k(1+\gamma)})^{\overline L-2n}+\overline L G_p
\end{IEEEeqnarray}
where
\begin{IEEEeqnarray}{l}
G_p:=\\
\sum_{k=\bar k(1+\gamma)}^{N}\!\! \!\binom{N}{k} \left(\frac{\overline k}{N}\right)^k \left(1-\frac{\overline k}{N}\right)^{N-k}\!\!\!\!\!\!\!\!\left(1-(1-p)^{k}\right)^{\overline L-2n}.\nonumber
\end{IEEEeqnarray}
It is now easy to verify that 
in the limit $p\to 0$:
\begin{equation}
G_p\leq \exp[-\gamma^2p^{-1}\log 2/2]+o(\exp[-\gamma^2p^{-1}\log 2/2])
\label{dpl}
\end{equation}
and by plugging \eqref{dpl} into \eqref{Cp3} the proof is completed.

\end{IEEEproof}

\section{Upper bounds on $\overline{T}(N,p)$ for $\beta\in[1/2,1)$
via Poisson-Poisson graphs}
\label{upper2}

In this section we prove Theorem \ref{teo:classbetal}. This allows to complete the proof of Theorem \ref{teo:betal}
which establishes upper and lower bounds on $\overline T/(Np|\log p|)$ when $p=1/N^{\beta}$ and $1/2\leq \beta<1$. 

\vspace{0.2 cm}

\IEEEproof
[Proof of Theorem \ref{teo:betal}]
The proof follows immediately from Theorem \ref{th:lower} and Theorem 
 \ref{teo:classbetal}. 

\endproof

\vspace{0.2 cm}

Furthermore Theorem \ref{teo:classbetal}  allows
 to identify a class of algorithms over
which the upper bound is attained. 

\begin{IEEEproof}[Proof of Theorem \ref{teo:classbetal}]
Consider the class of {\sl Poisson-Poisson distributions} on bipartite graphs defined in (\ref{poipoi}) with $N$ variable nodes, $M$ test nodes,  a mean number of tests  per variable equal to $L$
and a mean number of variables per test equal to $K=NL/M$, $P^{{\cal{P}}-{\cal{P}}}_{N,M,L}$.
From (\ref{aiha}), performing first the average with respect to the Poisson-Poisson distribution in which 
the  $c_{ia}$ variables are iid, the mean
number of undetected zeros can be written as:
\begin{IEEEeqnarray}{l}
\label{lastone}
\sum_{C_{N,M}}P^{{\cal{P}}-{\cal{P}}}_{N,M,L}(C_{N,M})\sum_X\mu_p(x)|{\cal{U}}_0|=\\
N \sum_X\mu_p(x)
 (1-x_1)
\left[1-\frac{K}{N} \prod_{j=2}^N\left(1-x_j\frac{K}{N}\right)\right]^M.
\nonumber
\end{IEEEeqnarray}

Denoting by $r$ the number of indices $j$ such that $x_j=1$, this gives:
\begin{IEEEeqnarray}{l}
\label{lasttwo}
\sum_{C_{N,M}}P^{{\cal{P}}-{\cal{P}}}_{N,M,L}(C_{N,M})\sum_X\mu_p(x)|{\cal{U}}_0|=\\
N \sum_{r=0}^{N-1}\binom{N-1}{r}p^rq^{N-r}\left(1 
-\frac{K}{N}(1-\frac{K}{N})^r\right)^M\nonumber
\end{IEEEeqnarray}
where we recall that $q:=(1-p)$.\\
Let $\gamma:=p/|\log p|$, then
\begin{IEEEeqnarray}{l}
\label{lastthree}
\sum_{r=N(p+\gamma)}^{N-1}\binom{N-1}{r} p^r(1-p)^{N-r}\\
<\exp[-N\gamma^2p^{-1}/2]+o(\exp[-N\gamma^2p^{-1}/2]).\nonumber
\end{IEEEeqnarray}
By using definition \eqref{meantest} and the above equations
\eqref{lastone}, \eqref{lasttwo} and \eqref{lastthree} we get
\begin{IEEEeqnarray}{l}
\lim_{N\to\infty\vert\beta}\sum_{C_{N,M}}P^{{\cal{P}}-{\cal{P}}}_{N,M,L}(C_{N,M})\frac{T_{M,C_{N,M},p}}{Np|\log p|}\leq\nonumber\\
\lim_{N\to\infty\vert\beta}\!{(p|\log p|)^{-1}}\!\left[\frac{M}{N}+\left(1
-\frac{K}{N}(1-\frac{K}{N})^{Np+N\gamma}\right)^M\right.\nonumber\\
\left.+N^{-1}\exp(-3Np|\log p|^{-2}/2)+p
\right]=\nonumber\\
\lim_{N\to\infty\vert\beta}\left[\frac{M}{Np|\log p|}+\frac{\left(1
-\frac{K}{N}(1-\frac{K}{N})^{Np+N\gamma}\right)^M}{p|\log p|}\right].\nonumber\\
\end{IEEEeqnarray}

By minimizing the last expression on $M$ and $K$ we find that the optimal value 
is taken on $M=epN|\log p|+o(Np|\log p|)=\widetilde M+o(\widetilde M)$
and $K=1/p+o(1/p)= N\widetilde L/\widetilde M+o(N\widetilde L/\widetilde M)$,
where $\widetilde L$ and $\widetilde M$ have been defined in \eqref{tildel} and \eqref{tildem}. Furthermore
\begin{equation}
\lim_{N\to \infty\vert \beta}
\sum_{C_{N,M}}P^{{\cal{P}}-{\cal{P}}}_{N,\widetilde M,\widetilde L}(C_{N,M})\frac{\overline{T}(N,p)}{Np|\log p|} \leq e
\end{equation}
is easily verified, thus completing the proof of Theorem \ref{teo:betal}.
\end{IEEEproof}
Remark \ref{byside} can  be proven along the same lines.

\section*{acknowledgment}

We thank Gregory Sorkin and Irina Rish for interesting discussions
which stimulated our interest in the group testing problem

\bibliographystyle{IEEEtran}
\bibliography{bibliGT}

\begin{thebibliography}{10}
\providecommand{\url}[1]{#1}
\csname url@samestyle\endcsname
\providecommand{\newblock}{\relax}
\providecommand{\bibinfo}[2]{#2}
\providecommand{\BIBentrySTDinterwordspacing}{\spaceskip=0pt\relax}
\providecommand{\BIBentryALTinterwordstretchfactor}{4}
\providecommand{\BIBentryALTinterwordspacing}{\spaceskip=\fontdimen2\font plus
\BIBentryALTinterwordstretchfactor\fontdimen3\font minus
  \fontdimen4\font\relax}
\providecommand{\BIBforeignlanguage}[2]{{%
\expandafter\ifx\csname l@#1\endcsname\relax
\typeout{** WARNING: IEEEtran.bst: No hyphenation pattern has been}%
\typeout{** loaded for the language `#1'. Using the pattern for}%
\typeout{** the default language instead.}%
\else
\language=\csname l@#1\endcsname
\fi
#2}}
\providecommand{\BIBdecl}{\relax}
\BIBdecl

\bibitem{Dorfman}
R.Dorfman, ``The detection of defective members of large poulations,''
  \emph{Ann.Math.Statist.}, vol.~14, pp. 436--440, 1943.

\bibitem{Zenios}
S.A.Zenios and L.M.Wein, ``Pooled testing for hiv prevalence estimation:
  Exploiting the dilution effect,'' \emph{Stat.Med.}, vol.~17, pp. 1447--1467,
  1998.

\bibitem{clone1}
B.~E.Barillot and D.Cohen, ``Theoretical analysis of library screening using
  and $n$-dimensional pooling strategy,'' \emph{Nuc.Acids Res}, vol.~19, pp.
  6241--6247, 1991.

\bibitem{clone2}
E.~D. C. N.~R. W.J.Bruno, D.J.Baldings and D.C.Torney, ``Design of efficient
  pooling experiments,'' \emph{Genomics}, vol.~26, pp. 21--30, 1995.

\bibitem{sequencing1}
D.Margaritis and S.Skiena, ``Reconstructing strings from substrings in
  rounds,'' \emph{Proc.Found.Comput.Sci.}, pp. 613--620, 1995.

\bibitem{sequencing2}
P.A.Pevzner and R.Lipshutz, ``Toward dna sequencing chips,'' in \emph{Proc.19th
  Int.conf.Math.Found.Comput.Sci.}, Lecture notes on computer sciences, 1994,
  pp. 143--158.

\bibitem{control}
M.Sobel and P.A.Groll, ``Group testing to eliminame efficiently all defectives
  in a binomial sample,'' \emph{Bell System tech. J.}, vol.~28, pp. 1179--1252,
  1959.

\bibitem{searching}
W.H.Kautz and R.C.Singleton, ``Nonrandom binary superimoposed codes,''
  \emph{IEEE Trans.on Information Th.}, vol.~10, pp. 363--377, 1964.

\bibitem{compression}
E.H.Hong and R.E.Ladner, ``Group testing for image compression,'' \emph{IEEE
  Trans.on Image Proc.}, vol.~11, pp. 901--911, 2002.

\bibitem{sensor}
Y.W.Hong and A.Scaglione, ``On multiple access for distributed dependent
  sensors: A content-based froup testing approach,'' \emph{Proc. IEEE
  Inf.Theory Workshop}, pp. 298--303, Oct 24-29 2004.

\bibitem{book}
D.Z.Du and F.K.Hwang, \emph{Combinatorial Group Testing and its
  Applications}.\hskip 1em plus 0.5em minus 0.4em\relax Singapore: World
  Scientific, 2000.

\bibitem{review1}
E.~D.J.Balding, W.J.Bruno and D.C.Torney, \emph{A comparative survey of
  nonadaptive pooling designs}.\hskip 1em plus 0.5em minus 0.4em\relax New
  York: T.S Speed and M.Waterman Eds., Springer Verlag, 1996, pp.133-154.

\bibitem{Knill}
E.Knill, ``Lower bounds for identifying subset members with subset queries,''
  in \emph{Proc.6th Ann.ACM-SIAM Symp.discr.Algorithms}, San Francisco, Jan.
  1995, pp. 369--377.

\bibitem{Berger}
T.Berger and V.I.Levenshtein, ``Asymptotic efficiency of two-stage disjunctive
  testing,'' \emph{IEEE Trans. on Inf.Th.}, vol.~48, pp. 1741--1749, 2002.

\bibitem{Vaccaro}
L.~A.De~Bonis and U.Vaccaro, ``Optimal two-stage algorithms for group testing
  problems,'' \emph{SIAM Journal on Computing}, vol.~34, pp. 1253--1270, 2005.

\bibitem{Berger2}
T.Berger and V.I.Levenshtein, ``A universal bound for a covering in regular
  posets and its application to pool testing,'' \emph{Discrete Applied
  Mathematics}, vol. 128, pp. 11--26, 2003.

\bibitem{LuMoura}
J.Lu and J.M.F.Moura, ``Structured ldpc codes for high-density recording: large
  girth and low error flow,'' \emph{IEEE Trans. on Magnetics}, vol.~42, pp.
  208--213, 2006.

\bibitem{FKG}
P.~C.M.Fortuin and J.Ginibre, ``Correlation inequalities on some partially
  ordered sets,'' \emph{Comm.Math.Phys.}, vol.~22, pp. 89--103, 1971.

\end{thebibliography}

\end{document}